\documentclass[12pt]{article} 

\usepackage{hyperref}
\usepackage{url}
\usepackage{setspace}
\usepackage{amsmath,amssymb,amscd,braket}
\usepackage{amsfonts}
\usepackage{color}
\usepackage{graphicx}
\usepackage{cite}

 \newcommand{\bea}{\begin{eqnarray}}
\newcommand{\eea}{\end{eqnarray}}
\newcommand{\be}{\begin{equation}}
\newcommand{\ee}{\end{equation}}
\newcommand{\ba}{\begin{align}}
\newcommand{\ea}{\end{align}}

\newcommand{\Tr}{{\rm {Tr}}}

\newcommand\rref[1]{(\ref{#1})}

\newlength{\slength}
\renewcommand{\title}[1]{\vbox{\center\LARGE{#1}}\vspace{5mm}}
\renewcommand{\author}[1]{\vbox{\center#1}\vspace{5mm}}
\newcommand{\address}[1]{\vbox{\center\footnotesize\em#1}}
\newcommand{\email}[1]{\vbox{\center\footnotesize\tt#1}\vspace{5mm}}

\numberwithin{equation}{section}

\begin{document}

\begin{titlepage}

\begin{center}

\hfill \\
\hfill \\
\vskip 1cm

\title{Euclidean Wormholes and Gravitational States}

\author{Alexandre Belin
}

\address{
Dipartimento di Fisica, Universit\`a di Milano - Bicocca \\
I-20126 Milano, Italy

\vspace{1em}
INFN, sezione di Milano-Bicocca, I-20126 Milano, Italy
}

\email{alexandre.belin@unimib.it}

\end{center}

\abstract{ Euclidean wormholes are known to encode important non-perturbative effects in the physics of quantum black holes. In this paper, we discuss the slicing of Euclidean wormholes along a time-reflection symmetric slice which treats half of the Euclidean geometry as a gravitational machinery to produce a semi-classical state. This type of state preparation is different from Hartle-Hawking states prepared with the CFT path integral, such as the thermofield-double state. Nevertheless, the two different types of  states have order one overlaps provided the gravitational data agrees on the initial data slice. This raises an interesting puzzle: one can easily construct an infinite family of semi-classical states that have order one overlap with the thermofield double state, while having a very different Euclidean preparation. We provide a microscopic description of wormhole states in the dual CFT and reformulate the factorization puzzle in the language of entanglement and the Hilbert space.

}

\vfill

\end{titlepage}

\eject

\tableofcontents

\section{Introduction}

Euclidean wormholes have continuously challenged our understanding of holography and the AdS/CFT correspondence, ever since the discovery of the factorization puzzle \cite{Maldacena:2004rf}: the product of CFT partition functions on disconnected Euclidean manifolds does not factorize when computed in the gravitational theory, violating a fundamental property of quantum mechanics. For several years, this led to the belief that these geometries should not be taken seriously, either because of instabilities or because they should be excluded by hand from the gravity path integral. More recently, it has been understood that these geometries are in fact encoding important non-perturbative effects in the physics of quantum black holes. In particular, they provide a partial resolution of the black hole information paradox \cite{Maldacena:2001kr}, by preventing bulk calculations from violating the unitarity of quantum mechanics \cite{Saad:2018bqo,Saad:2019lba}. While these developments do not resolve the factorization puzzle, they underline how important it is to listen to the messages Euclidean wormholes are telling us, and it is clear that a better understanding of these geometries is central to address some of the deepest puzzles in quantum gravity.

\begin{figure}
\centering
  \includegraphics[width=.8\linewidth]{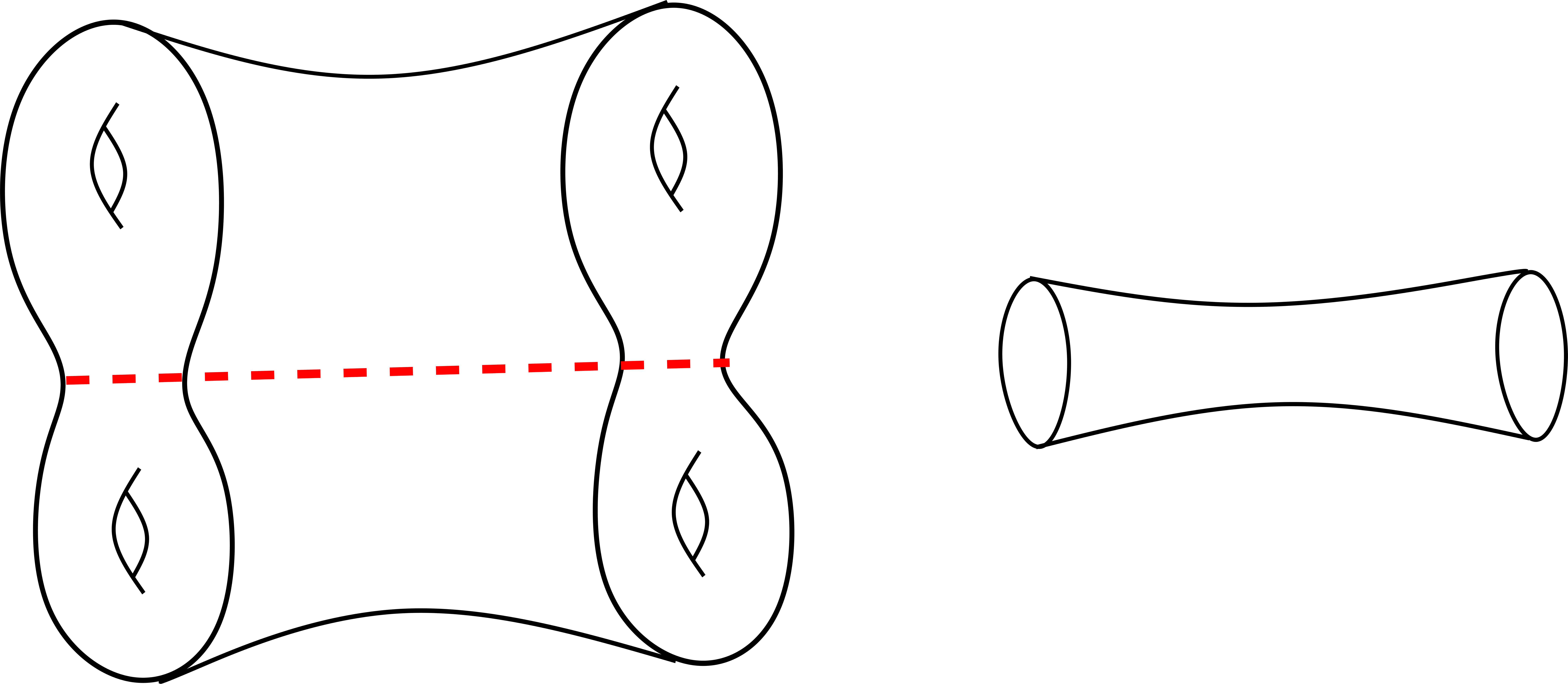}
\caption{On the left, a slicing of the genus-two wormhole on a time-reflection symmetric slice. On the right, the geometry of that slice, with topology circle times an interval.}
  \label{cutting}
\end{figure}

 In this paper, we explore a different facet of Euclidean wormholes: we view these geometries as a gravitational machinery to prepare semi-classical states. The idea is to take a Euclidean wormhole, slice it along a $\mathbb{Z}_2$-symmetric slice, and analytically continue the gravitational data on that slice to obtain Lorentzian initial data.\footnote{It is important to emphasize that the slicing we do here is different than the one of \cite{Antonini:2022blk,Antonini:2022ptt,Betzios:2024oli}, which wanted to prepare initial data in a closed universe. Here, our initial data slice will always touch the asymptotically AdS boundaries.} We summarize this in Fig. \ref{cutting}. This is similar to what we do in quantum field theory when we use the Euclidean path integral to prepare states, although as we will see, the preparation of semi-classical states through wormholes cannot be associated to any Euclidean path integral preparation in the dual CFT. Classically, the Euclidean geometry is an on-shell way to prepare Lorentzian initial data, and thus it is natural to think that there should be some semi-classical state $\ket{\psi_{\text{WH}}}$ corresponding to this preparation. 
 
This idea was first discussed in \cite{Hubeny:2007xt}.  Since the initial data geometry has two asymptotically AdS boundaries and a connected spatial slice, it is natural to think this state is an entangled state living in the tensor product of two CFT Hilbert spaces
\be
\ket{\psi_{\text{WH}}} \in \mathcal{H}_L \otimes \mathcal{H}_R \,.
\ee
The first main result of this paper will be to propose a microscopic dual for this wormhole state. We propose the state to be given by
\be \label{microintro}
 \ket{\psi_{\text{WH}}}  = \sum_{i,j\neq 1} |C_{iij} |^2 \mathbb{O}_j ^L \bar{\mathbb{O}}_j^R \ket{0} \otimes \ket{0} \,.
\ee
Here, $C_{iij}$ are CFT OPE coefficients. $\mathbb{O}$ represents the half-genus two OPE block corresponding to some Virasoro primary operator $O_j$. This means that it corresponds to the action of the primary operator $O_j$, along with all of its descendants plus the descendants of the primary operator that runs in the loop of the half genus-two surface. The two-point function of such an OPE block in the vacuum produces a genus-2 conformal block, and more details will be given in section \ref{sec:microstate}. The operator $\bar{\mathbb{O}}$ is the operator CPT conjugate to $\mathbb{O}$. The state \rref{microintro} is currently unnormalized, and as we will see, its norm gives the partition function of 3D gravity on the Euclidean wormhole geometry.

At the level of initial data, there is another state that seems very similar to the one described above, the thermofield double state
\be
\ket{TFD} = \sum_{i} e^{-\beta E_i/2} \ket{i} \otimes \ket{\bar{i}} \,.
\ee
This state is also an entangled state in the Hilbert space of two CFTs, and above the Hawking-Page transition, it describes the eternal two-sided AdS black hole \cite{Maldacena:2001kr}. Its initial data surface in the bulk is a cylinder connecting the two boundaries, just like the wormhole state. The preparation of the thermofield-double state is done with the Euclidean path integral in the CFT, which as we will review, is a standard way to construct semi-classical states \cite{Skenderis:2008dh,Botta-Cantcheff:2015sav,Marolf:2017kvq,Belin:2018fxe}. It is very natural to ask what the overlap of the two different states are and this is the second main result of the paper: we will show that the two states have a large overlap in the semi-classical limit
\be
\frac{\braket{\psi_{\text{WH}}| \text{TFD}}}{\sqrt{\braket{\psi_{\text{WH}}|\psi_{\text{WH}}}\braket{\text{TFD}| \text{TFD}}}} \approx 1 \,.
\ee
This can be understood the most naturally from the bulk theory. The initial data for both states is the same, and thus we can cut half of the Euclidean wormhole and half of the Euclidean BTZ black hole, and smoothly glue them together. To leading order, the overlap is given by the exponential of the on-shell action of the glued geometry, suitably normalized by the overlaps of the states with themselves. Since the action is a local functional and the gluing is smooth, the overlap gives one.

This presents an interesting puzzle: the same initial data can be prepared by two very different Euclidean geometries, and one can thus find a large (infinite, since there are continuous moduli that can be varied) number of states which are not the thermofield-double state, but have order one overlap with it in the large $c$ limit\footnote{A similar situation was encountered in JT gravity \cite{Anderson:2020vwi}. Another interesting parallel is the TFD double state in two decoupled SYK models and the ground state of the Maldacena-Qi Hamiltonian \cite{Maldacena:2018lmt}. Both states also appear to have order one overlap in the large $N$ limit. We thank an anonymous referee for bringing this point to our attention.}. This is a sort of inverse problem for the over-completeness of states living in the gravitational EFT, and the resulting consequence which is that of null states. There, this is a sharp tension with quantum mechanics, as we can only fit so many states in a finite size Hilbert space. This is the black hole information paradox. Here, there is no tension with quantum mechanics, but it is still surprising to find so many semi-classically different ways to prepare states that are very close in the Hilbert space. This underlines the ill-posedness of the Euclidean initial value problem, as already shown at the perturbative level in \cite{Belin:2020zjb}, and makes it clear that one cannot unambiguously find \textit{the} state dual to a semi-classical geometry.

The rest of the paper is organized as follows: in section \ref{sec:coherentstates}, we review the standard construction of semi-classical (or generalized Hartle-Hawking) states in AdS/CFT using the CFT path integral. In section \ref{sec:gluing}, we present the slicing of the Euclidean wormhole and discuss its classical overlap with the thermofield double state. In section \ref{sec:microstate}, we present the microscopic construction of the state and discuss its properties. We conclude with some comments and open questions in section \ref{sec:discussion}.

\section{Semi-classical states in AdS/CFT \label{sec:coherentstates}}

One of the slogans we often hear in the AdS/CFT correspondence is that states are dual to geometries. For example, the vacuum  is dual to the vacuum AdS geometry and a thermal state (at high enough temperatures) is dual to a black hole. However, this slogan needs to be refined. First of all, not all CFT states are dual to semi-classical geometries, but rather only very special ones. Second, if we are talking about a Lorentzian asymptotically AdS spacetime, this does not correspond to a state in the CFT, but rather to the entire time-history of a state: its evolution with the CFT Hamiltonian. 

A semi-classical state should be described by the data at one moment in time. In the bulk, at the classical level, this data is simply the data on a Cauchy slice needed for the initial value problem of general relativity. This means we need the induced metric on a Cauchy slice, as well as the first normal derivatives (i.e. the extrinsic curvature). A semi-classical state is thus described by the data
\be
\ket{h_{ab},K_{ab}} \,,
\ee
and similarly for matter fields if they are present in the gravitational theory. So far, this is similar to what is done in classical field theory, but gravity is a gauge theory and the Hamiltonian constraint presents an extra challenge:\footnote{We will not discuss the momentum constraints as those can be dealt with similarly to ordinary gauge theories.} the specification of the choice of Cauchy slice is a gauge redundancy. This will not be important in this paper, but we briefly discuss it for completeness. Because of the Hamiltonian constraint, the dual of a CFT state is not the data on a Cauchy slice, but rather the entire Wheeler-de Witt patch anchored at a boundary time. The Wheeler-de Witt patch is foliated by Cauchy slices, and one can "evolve" from one slice to another using the constraint equation. This is best done using the prescription of York \cite{PhysRevLett.28.1082} (see \cite{Belin:2018bpg} for a  review in the context of AdS/CFT). One separates the induced metric into a conformal metric and a scale factor, and separates the extrinsic curvature into a traceless and a trace part. One can take the trace of $K=h^{\mu\nu} K_{\mu\nu}$ to be constant on the different slices (constant mean curvature slices) and $K$ then serves as a time inside the Wheeler de-Witt patch, labelling the choices of Cauchy slices. The conformal factor can then be solved for by knowing the rest of the initial data along with the value of $K$. 

Note that the data on a Cauchy slice contains twice the amount of data needed to describe a quantum state. In analogy with quantum mechanics, the wave-function should be a function of either $x$ or $p$, but not both. This is not a problem, since these states should be thought of as coherent states of the quantum gravity theory: the most classical states allowed by quantum mechanics. For the SHO, the coherent states are labelled by a choice of a complex parameter $\alpha$ whose real and imaginary parts fixes the expectation values of $x$ and $p$. One should view the states $\ket{h_{ab},K_{ab}}$ in the same way. These states are as classical as possible, meaning that the variance of these quantities is small and controlled by $\hbar$, whose role in gravity is played by $G_N$.

The important question is to understand which CFT states describe the semi-classical states $\ket{h_{ab},K_{ab}}$. Much work towards answering this question was done over the last decade, and we now have a good picture of what these states are: they are states prepared by the Euclidean path integral with sources turned on for single-trace operators \cite{Botta-Cantcheff:2015sav,Marolf:2017kvq,Belin:2018fxe}. These are often referred to as generalized Hartle-Hawking states. We will refer to these states as
\be
\ket{\lambda(x)} \,.
\ee
The wave-function of these states is given by
\be
\braket{\varphi_0|\lambda(x)}=\int_{\varphi(t_E=0)=\varphi_0} \mathcal{D}\varphi e^{-S_{\text{CFT}}+\int_{t_E<0} \sum_i \lambda^i(x) O^i(x)} \,,
\ee
where the sum over $i$ labels the different single-trace operators in the CFT. This can be done on any topology in the CFT, but is easiest to think about on the sphere. We do the Euclidean path integral on the southern hemisphere, and turn on sources for single-trace operators. Note that in general, the sources are complex here. These sources should be viewed as the generalization of the parameter $\alpha$ in the SHO coherent states. The sources need to vanish sufficiently fast as $t_E\to 0$ to guarantee that the state is normalizable, or viewed differently, to guarantee that this preparation gives an excited state in the original CFT Hilbert space, rather than a state in the Hilbert space of a deformed theory.

The mapping between the initial data and the source $\lambda(x)$ is non-trivial, and is obtained by solving Einstein's equations using the prescription of Skenderis and Van Rees \cite{Skenderis:2008dh}. We consider the overlap of the state with itself
\be
\braket{\lambda(x) | \lambda(x)} \,.
\ee
The bra state is prepared by the Euclidean path integral on the northern hemisphere with conjugate sources. After solving Einstein's equations in the Euclidean setup with these complex sources, we look for a bulk slice where the induced metric is real and the normal derivatives are purely imaginary. Such a slice should exist by $\mathbb{Z}_2+\mathbb{C}$ invariance of the boundary conditions. On that slice, one analytically continues the initial data to Lorentzian signature and we have a well-defined initial value problem with real initial data, giving a real Lorentzian spacetime. 

A couple comments are important. First note that any state in the CFT can be written as
\be
\ket{\psi} = \sum_i c_i \ket{E_i} \,,
\ee
where $E_i$ are the energy eigenstates on the sphere which are in one-to-one correspondence with the local operators of the CFT through the state operator correspondence. With infinite power, one should in principle be able to understand how to describe the states $\ket{\lambda(x)}$ in this language, but this would require to first know the exact spectrum of the CFT (including that of black hole microstates) and moreover to track the exact values of the coefficients $c_i$ by expanding the exponentials, using the OPE and by performing the integrals. Needless to say that this is hopeless with semi-classical methods. The advantage of the Euclidean path integral description of the state, is that instead of microscopically describing the state, we can semi-classically describe \textit{the preparation} of the state. And it is this preparation that has a nice mapping to semi-classical variables of the bulk theory.

Second, the map
\be
\ket{\lambda(x)}  \Longleftrightarrow \ket{h_{ab},K_{ab}} \,,
\ee
is actually subtle. The direction from $\lambda(x)$ into the bulk is best understood. The Euclidean boundary value problem is well-posed in the PDE sense, and the existence of a solution is even mathematically proven in some finite size regions of source space \cite{Anderson:2001pf,Hickling:2015tza,Fischetti:2016vfq}\footnote{Uniqueness is in general hopeless, as there can be multiple different topologies that contribute in the bulk. For example, if the boundary is a torus in $d=2$, then we know of infinitely many ways to fill the bulk smoothly, two of which are thermal AdS and the BTZ black hole. The initial data dual to a source should always be taken as the data coming from the leading saddle.}. The other direction is actually much more subtle. In the PDE sense, specifying initial data and looking for the source that prepares it at the boundary of AdS is ill-posed. This can be shown rigorously at the linearized level \cite{Belin:2020zjb}, and can be expected to extend at the non-linear level. In fact, the results of this paper will be a critical failure of this direction of the map, as we will find several Euclidean preparations corresponding to the same initial data.

Third, the initial data corresponding to a state is only relevant in the overlap of a state with itself. If we consider overlaps of different coherent states
\be \label{overlapHH}
\braket{\lambda'(x) | \lambda(x)} \,,
\ee
there is no sense in which the initial data corresponding to either $\lambda$ or $\lambda'$ is relevant to the geometry computing this overlap.\footnote{This statement is meant to hold if the sources are different by a finite amount. If the sources are perturbatively close, then more structure is present, as  is for example exploited in relating bulk and boundary symplectic forms \cite{Belin:2018fxe}.} In general, these overlaps are non-zero, but of size
\be
e^{-\frac{1}{G_N}} \,.
\ee
This can be understood on general grounds from the over-completeness of coherent states, and physically means that each semi-classical geometry has tails in its wave-function with support on other semi-classical geometries. These are non-perturbatively small, as the wave-function is highly peaked, but they are non-zero. This is true even if the topology of the initial data slice is different between the two coherent states \cite{Jafferis:2017tiu}.

Generalized Hartle-Hawking states have found many interesting applications in AdS/CFT, for example the derivation of Einstein's equation through entanglement \cite{Faulkner:2017tkh}, the equivalence between bulk and boundary symplectic forms \cite{Belin:2018fxe} or more recently the formulation of state-dressed operators in AdS/CFT \cite{Bahiru:2022oas,Bahiru:2023zlc}. They also include several generalization, for example by considering the addition of multi-trace sources \cite{Haehl:2019fjz}. We refer to these papers and references therein for more details, as the main properties useful for this paper are summarized above. We now turn to the description of the Lorentzian slicing of the wormhole, as a different way of gravitationally preparing semi-classical states.

\section{Slicing the Euclidean wormhole and the overlap with BTZ \label{sec:gluing}}

In this section, we will describe the slicing of the Maoz-Maldacena wormhole which produces a semi-classical state, upon analytic continuation to Lorentzian signature. We will also show that the overlap of this state with the thermofield double-state is one  in the semi-classical limit.

The metric of the wormhole is given by
\be
\frac{ds^2}{\ell^2}= d\rho^2+\cosh^2\rho d\Sigma_{g=2}^2 \,.
\ee
Here $d\Sigma_{g=2}^2$ is the constant negative curvature metric on a compact genus-two surface. The genus-2 surface can be obtained from the upper half plane by a quotient with respect to the appropriate discrete subgroup of the isometry group of the upper half plane. The three-geometry has three parameters encoded in the moduli of the genus-2 surface. We take the genus-2 surface to have a $\mathbb{Z}_2$ symmetry, with fixed point given by the cut shown in Fig. \ref{cutting}. Within this slice of moduli space, the genus-2 surface only has two independent moduli: one labelling the length of the handles (both are set to be the same by the $\mathbb{Z}_2$ symmetry), and one labelling the size of the neck connecting the two handles. The symmetry of the genus-2 surface extends into a symmetry of the full 3-geometry, and we can slice the 3-geometry on the $\mathbb{Z}_2$-symmetric slice. The metric of the three-manifold near that surface is
\be \label{metricnearcut}
ds^2\approx d\rho^2+\cosh^2 \rho (dr^2 + \cosh^2 r dx^2) \,,
\ee
with $r$ labels a coordinate of $\Sigma_{g=2}$ normal to the cut, while $x$ is the other coordinate which extends along the cut. Away from the cut, the metric is much more complicated since there is a handle, but locally near the cut, the metric takes this form. The coordinate $x$ is identified as
\be
x\sim x+ x_0 \,,
\ee
and $x_0$ represents one of the moduli of the genus-2 surface. The other modulus can obviously not be encoded in the metric near the cut. It is clear from the metric \rref{metricnearcut} that the geometry represents initial data given by 
\bea \label{kin}
K_{ab}=0 \,,
\eea
and $h_{ab}$ representing the two-metric
\be \label{hin}
\frac{ds^2}{\ell^2} = d\rho^2 + \cosh^2 \rho dx^2 \,.
\ee
It is interesting that the modulus of the handle living far away from the cut has absolutely no impact on the initial data. So we can change that modulus at will and still have the exact same state in the classical limit. We will come back to this below.

It is important to note that this is \textit{not} a Hartle-Hawking state. This state is not obtained by fixing the CFT boundary and performing the Euclidean path integral on it, which should be viewed as the AdS equivalent of the no-boundary prescription. This state is prepared differently, and at this stage should simply  be viewed as arising from doing the perturbative gravity path integral around some fixed on-shell saddle. Whether there is a non-perturbative and microscopic description of the state is unclear at this level (in contrary to Hartle-Hawking states which are manifestly microscopic since they correspond to a CFT path integral), but we will see in the next section that we actually \textit{can} find a microscopic definition of the wormhole state, even if it is not coming from a Euclidean path integral in the CFT.

The initial data \rref{kin} and \rref{hin} should be easily recognized as being the same as that of the (spinless) BTZ black hole. The one free modulus $x_0$ encodes the temperature of the black hole. It is then very natural to think that one can smoothly glue one half of the BTZ metric to one half of the Euclidean wormhole.

To see this more explicitly, consider the following coordinates for the BTZ black hole
 \be \label{BTZhyper}
 ds^2= d\rho^2 + \cosh^2 \rho (dr^2 + \cosh^2 r dx^2) \,.
 \ee
Again, $x$ is a periodic direction with period $x_0$. Otherwise this would just  be the metric on $\mathbb{H}^3$. These coordinates, along with identification of the $x$ coordinate, make explicit the fact that the BTZ geometry is a $\mathbb{Z}$ quotient of $\mathbb{H}^3$. $x_0$ is precisely the area of the horizon of the BTZ geometry.

These coordinates are hyperbolic coordinates for the BTZ black hole, and we briefly comment on some of their properties. The boundary is obtained by taking either $\rho\to \pm \infty$, or by taking $r\to\pm\infty$. The boundary is a torus, which one should think of as representing the overlap of the TFD state with itself. Each TFD state is obtained by doing the path integral on an annulus, a northern annulus for the bra and a southern annulus for the ket. The limit $\rho\to\infty$ gives the northern annulus, while $\rho\to-\infty$ gives the southern annulus. The limit $r\to\infty$ goes to the R circle of the initial data slice ($\rho=0$), while $r\to-\infty$ goes to the L circle.
 
Given these coordinates, it should now be clear that we can glue the two geometries smoothly. In fact, near the cutting surface the two coordinates \rref{metricnearcut} and \rref{BTZhyper} are the same! The fact that we can smoothly glue the two geometries implies that the overlap is one in the classical limit. Consider the normalized overlap

\be
\frac{\braket{\psi_{\text{WH}}| \text{TFD}}}{\sqrt{\braket{\psi_{\text{WH}}|\psi_{\text{WH}}}\braket{\text{TFD}| \text{TFD}}}} \approx \frac{e^{-S_{\text{on-shell}}[\text{glued geometry}]}}{e^{-S_{\text{on-shell}}[\text{WH}]/2-S_{\text{on-shell}}[\text{BTZ}]/2}} \,.
\ee
Because the Einstein-Hilbert action is local, and the BTZ and wormhole geometries have a $\mathbb{Z}_2$ symmetry, the BTZ/wormhole actions can be split into a contribution from the northern and southern components. Taking half the action is equivalent to taking exactly only the northern or southern component, which is precisely what the action of the glued geometries computes. The smooth gluing guarantees that no extra boundary terms arise at the gluing surface, and hence we obtain
\be
\frac{\braket{\psi_{\text{WH}}| \text{TFD}}}{\sqrt{\braket{\psi_{\text{WH}}|\psi_{\text{WH}}}\braket{\text{TFD}| \text{TFD}}}} \approx 1 \,.
\ee

\begin{figure}
\centering
  \includegraphics[width=.5\linewidth]{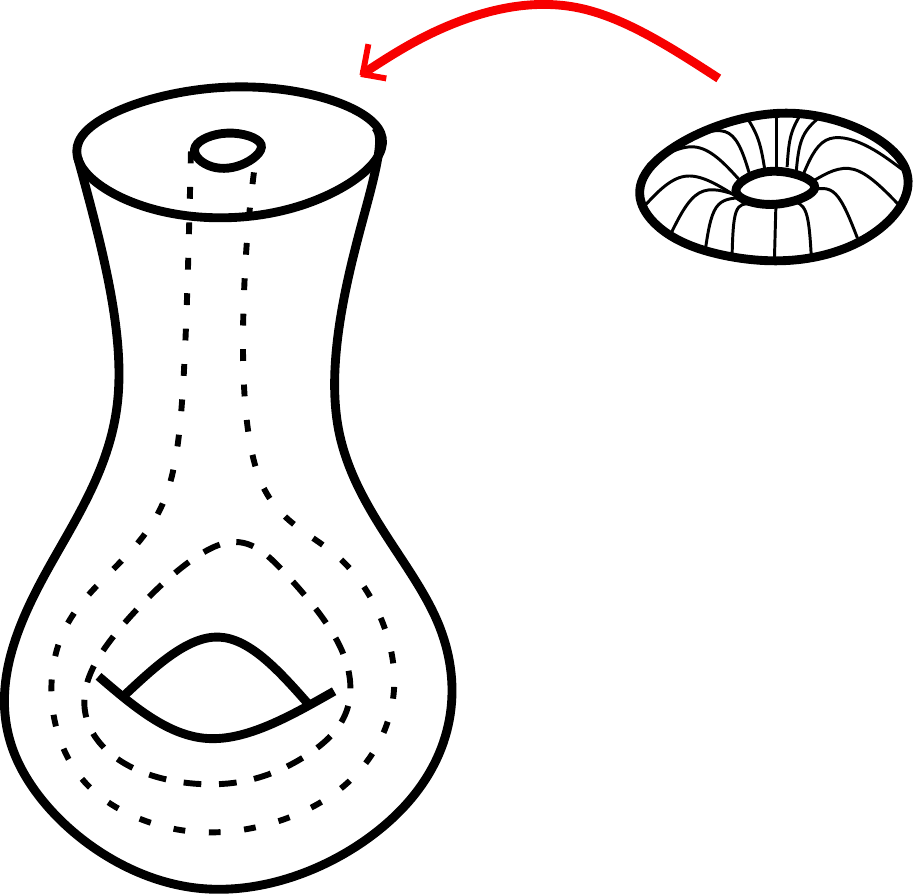}
\caption{The gluing of the two geometries. On the left, we have half the Euclidean wormhole geometry. The two asymptotically AdS boundaries are the solid black line and the dashed line. The three-geometry is the filling between these two surfaces. We see that the dashed line links the other handle in a non-trivial way. On the right, half the BTZ geometry (the bagel slicing). We see that we can smoothly glue it at the top of half the wormhole. The single boundary of half-BTZ (an annulus) then glues together the two half genus-2 surface, into a single boundary of genus 2.}
  \label{gluing}
\end{figure}

It is worthwhile to describe the geometry we obtained after gluing. The easiest representation for it is represented in Fig. \ref{gluing}. The idea is that we start by drawing the wormhole geometry as two genus-2 surfaces, one inside the other. The wormhole is the geometry that fills up the space between this two genus-2 surfaces. Then we cut it in half. On the other hand, the BTZ is a solid torus, and the appropriate cutting to obtain the BTZ geometry (rather than thermal AdS) is the "Bagel" slicing of the geometry. From these two representations, one can easily see that we can glue one half to the other. The outcome of this gluing is a geometry with a single AdS boundaries: both boundaries of the half genus-2 wormhole have been tied together by the half bagel of BTZ. The three-geometry represents a filling of this genus-2 surface, and hence is a genus-2 handle-body. Note that the knotting of the handle-body is non-trivial, as one of the handles is linked to the other. They are not linked together in the obvious way though, but rather one handle is linked to the dual cycle of the other handle.

One important comment is in order: because we are not discussing solely Hartle-Hawking states (the wormhole geometry is not specified by fixing boundary conditions for the gravitational path integral), there is no boundary condition that instructs us to compute this overlap. This is different from usual Hartle-Hawking type states prepared by the gravitational path integral with sources turned on, where the overlap \rref{overlapHH} would be computed by some fixed boundary conditions. Here, we just glue two geometries together and find a geometry with a single boundary, which can be viewed as some boundary condition. But it is not clear that the glued geometry is actually the dominant saddle corresponding to those boundary conditions. In principle, there could be another saddle with the same boundary conditions but with smaller action, and whose geometry would have nothing to do with the gluing of the half BTZ and half wormhole. We will revisit this question when we describe the state microscopically.

We can now see that there are an infinite number of states whose overlap with the BTZ gives one at leading order. First of all, there is the modulus of the handle of the wormhole that did not enter here, so any value of that modulus does not change the overlap. Second, it is easy to generalize this construction to include a wormhole with arbitrary number of handles (and hence also arbitrary number of moduli). We consider a very complicated Riemann surface with $n$ handles and restrict to the slice of moduli space that has the desired $\mathbb{Z}_2$ symmetry about a circle of the surface. Then, we slice. The geometry on the time-slice is again exactly the cylinder and all details of the Riemann surface have been washed out. For the same reasons as above, the overlap with the BTZ is one in the leading large $c$ limit. The overlap between the different Euclidean wormhole states is also one. This gives a puzzle.\footnote{We insist that this is not a paradox, as it does not violate any property of quantum mechanics.} An infinite number of extremely distinct Euclidean preparations give rise to many states which are in fact very close in the Hilbert space. We comment further on this in the discussion.

Before turning to the microscopic description of the state, we comment on the extent to which the states are different, and can be identified as different. It is only the bulk timeslice geometry that is the same. The preparations are very different, and as such, they will give rise to very different quantum states for the bulk fields. This means that there are ways to distinguish the two states. One way would be to compute the order one contribution to the entanglement entropy of the left (or right) CFT. Another would be to compute a left-right two-point function, even if the difference would not be visible in the geodesic approximation. These facts also resonate with the fact that while the leading term in the overlap is one at large $c$, it can have a potentially small geometric prefactor. We expand on this in the discussion.

\section{A microscopic description of the wormhole state\label{sec:microstate}}

We will now propose a microscopic dual for the state. The main ingredient we will need is some insight from 3D gravity, where it was realized in \cite{Belin:2020hea} (and expanded on in \cite{Chandra:2022bqq}) that Euclidean wormholes come from treating CFT OPE coefficients statistically, with a distribution that is to leading order a Gaussian.\footnote{The non-Gaussianities are in fact very important for the general picture of 3D gravity, and they have been discussed in \cite{Belin:2021ryy,Collier:2023fwi,Collier:2024mgv,Belin:2023efa,Jafferis:2024jkb,deBoer:2024mqg,Hartman:2025ula,Chandra:2025fef}. But they will play no role in this paper.}

\begin{figure}
\centering
  \includegraphics[width=.2\linewidth]{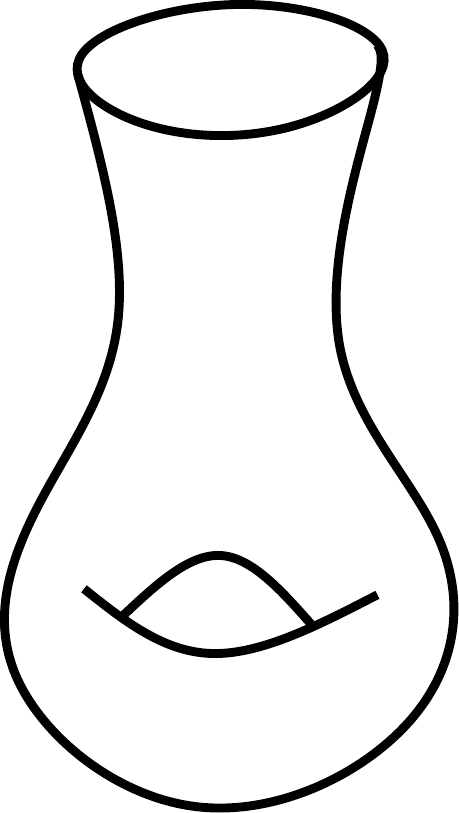}
\caption{The state $\ket{\psi_{\frac{1}{2}g=2}}$ is prepared by doing the Euclidean path integral on half of a genus-2 surface. This state has two parameters, related to the two moduli of the surface. One is related to the length of the handle, while the other is related to the length of the cylinder extending from it.}
  \label{halfgenus2}
\end{figure}

To find the wormhole state, we start by considering a state prepared by the Euclidean path integral on half a genus-2 surface, as represented in Fig. \ref{halfgenus2}, and related to the torus operator discussed in \cite{Marolf:2017vsk}. We will call this state
\be
\ket{\psi_{\frac{1}{2}g=2}} \,.
\ee

In terms of CFT data, this state is given by
\be
\ket{\psi_{\frac{1}{2}g=2}} = \sum_{i,j} C_{iij} \mathbb{O}_j \ket{0} \,.
\ee
Here, $C_{iij}$ are the OPE coefficients of the CFT, and the sum over $i$ and $j$ runs over primary operators. The object $\mathbb{O}_j$ is what we call a half genus-2 OPE block. This means its starting point is the insertion of the primary operator $O_j$, but it is dressed with the contributions of all descendants, both of the external states $j$, but also the internal states $i$ that run in the loop. It implicitly depends on the two moduli of the half genus-2 surface, as the contribution of all operators come with coefficients depending on the moduli. The two-point function of the half genus-2 OPE block in the vacuum gives a genus-2 Virasoro block (or rather a square thereof, if we take into account left and right movers). We have
\be
\braket{0 | \mathbb{O}_j \mathbb{O}_k | 0 } = \delta_{j,k} | \mathcal{F}_{g=2}(h_j,h_{\text{internal}})|^2 \,.
\ee
We now consider the double-copy of the CFT Hilbert space (we will label the two factors L and R), and consider the product of two of these states. The product state reads
\be \label{prodstate}
\ket{\psi_{\frac{1}{2}g=2}}_L \otimes \ket{\psi_{\frac{1}{2}g=2}}_R = \sum_{i,j,k;l} C_{iij} C_{kkl} \mathbb{O}_j^L \mathbb{O}^R_l\ket{0}_L \otimes \ket{0}_R \,.
\ee

If we took the overlap of this state with itself, we would obtain two disconnected genus-2 partition functions. We now use the idea that the wormhole comes from the connected Gaussian contraction between the OPE coefficients of the two distinct partition functions. In practice, this implements a diagonal projection on the sums over primary operators. This hints that we should be doing a diagonal projection to obtain the wormhole state. Doing a diagonal projection on unentangled states is a natural way to entangle them, which in the bulk should connect the topology of the spatial slice. This is exactly what will happen. Our conjecture is that the wormhole state is given by\footnote{Note that a similar type of diagonal projection for operators has appeared in \cite{Betzios:2021fnm,Betzios:2023obs}. It would be interesting to understand the connection better.}
\be \label{ansatzwh}
\ket{\psi_{\text{WH}}}= \sum_{i,j \neq 1} |C_{iij}|^2  \mathbb{O}_j^L \bar{\mathbb{O}}^R_j\ket{0}_L \otimes \ket{0}_R \,.
\ee
Here the notation $\bar{\mathbb{O}}$ means that we take the CPT conjugate operator. This is implemented by the nature of the diagonal projection in the Gaussian ensemble \cite{Chandra:2022bqq} and the way operators are paired. The identity is also removed from the sum, since there is no non-trivial contraction for the identity operator. This is an entangled state in the tensor product of two CFTs.\footnote{Note that this is in some sense the maximally entangled state built from the two previous states we had. We could also have considered another state where we only pair the state $j$ and $l$ in \rref{prodstate}, and we would also have produced an entangled state. The intuition from the wormhole partition function tells us to pair all the states, but it would be interesting to understand whether this partially entangled state has a gravitational description.}

We will now give support for the conjecture that this state is dual to the wormhole state in the bulk. First, note that the state is not normalized. The norm of the state is precisely the wormhole partition function in gravity. We have
\be \label{overlapwithitself}
\braket{\psi_{\text{WH}}|\psi_{\text{WH}}} =  \sum_{i,j,k,l \neq 1 } |C_{iij}|^2  |C_{kkl}|^2 \braket{\mathbb{O}_l \mathbb{O}_j}_L \braket{\bar{\mathbb{O}_l} \bar{\mathbb{O}_j}}_R = \sum_{i,j,k \neq 1}  |C_{iij}|^2  |C_{kkj}|^2 |\mathcal{F}_{g=2}|^4 \equiv Z_{\text{WH}} \,,
\ee
which is exactly the genus-2 wormhole partition function, in the dumbbell channel \cite{Belin:2020hea,Chandra:2022bqq}.\footnote{One should be careful about the range of the indices that run in the sum. The identity is not summed over here, which follows from \rref{ansatzwh}. Had it been included, we would not have had the wormhole contribution, but rather the sum of the wormhole contribution together with a handlebody contribution, and the handlebody contribution would have been larger, overshadowing the wormhole piece.}

\subsection{The overlap with the TFD state}

We can also compute the overlap between our conjectured wormhole state and the thermofield double state. We have
\be
\braket{\text{TFD}|\psi_{\text{WH}}}= \sum_{k} e^{-\beta E_k/2} \bra{E_k} \otimes \bra{\bar{E}_k} \sum_{i,j} |C_{iij}|^2  \mathbb{O}_j^L \bar{\mathbb{O}}^R_j\ket{0}_L \otimes \ket{0}_R \,.
\ee
The sum over $k$ will be projected down to match the $j$ energy eigenstates due to orthonormality. The role of the factor of $e^{-\beta E_k /2}$ will be to "glue" the two half genus-2 OPE blocks into a genus-2 conformal block. The modulus corresponding to the neck of the cylinder will be altered in this process. Roughly, if we think of $\beta_{\text{WH}}$ as the length of the cylinder in the half genus-2 state, the length of the cylinder of the genus-2 surface obtained after gluing the two OPE blocks together will be
\be
\beta_{g=2} \sim 2 \beta_{\text{WH}} + \frac{\beta}{2} \,.
\ee
Note that the way we obtained a genus-2 conformal block is different compared to the overlap of the wormhole state with itself. In that case, the half genus-2 OPE block in the one copy of the ket got paired up with another half genus-2 OPE block in one copy of the bra. And similarly for the other copy. Here, the thermofield double state glues together the two copies in the ket state into a single genus-2 OPE block. All in all, this gives
\be \label{genus2parttilde}
\braket{\text{TFD}|\psi_{\text{WH}}}= \sum_{i,j} |C_{iij}|^2  |\tilde{\mathcal{F}}|^2 \equiv \tilde{Z}_{g=2} \,.
\ee
We have denoted the conformal block as $\tilde{\mathcal{F}}$ to emphasize that one of the moduli of that genus-2 surface are different than those that appeared in \rref{overlapwithitself}. We recognize here a genus-2 partition function, expanded in the dumbbell channel. This resonates with the fact that the geometric gluing of the two geometries (BTZ and Euclidean wormhole) performed in the bulk gave rise to a genus-2 handlebody. It is natural to expect that the handlebody found in the bulk is the dominant saddle-point geometry with the moduli prescribed by the genus-2 surface of \rref{genus2parttilde}, but it is not guaranteed from first principles at this stage.

In total, for the normalized overlap, we have
\be
\braket{\text{TFD}|\psi_{\text{WH}}}= \frac{\tilde{Z}_{g=2} }{\sqrt{Z(\beta) Z_{\text{WH}}} }\,.
\ee
We emphasize that this is a CFT expression, as all partition functions appearing here are defined microscopically in the CFT. Given this expression, it is perhaps less surprising that the overlap can be one to leading order at large $c$. This is a ratio of particular partition functions, and the bulk leads us to conclude that moduli of the different partition functions have simply been tuned to give an overlap which is approximately one. We will not evaluate this ratio here and explicitly check that the ratio gives 1, as it is hard to evaluate a genus-2 partition function in closed form. Indeed, one needs a good knowledge of the associated genus-2 blocks and one needs to pick coordinates on moduli space. But this could be done numerically at large $c$, using the prescription of \cite{Maxfield:2016mwh}. 

In any case, we see here that there is no sharp tension with quantum mechanics. We are finding states that have overlaps close to one, achieved by a careful matching of moduli, but this does not contradict any fundamental principle. It is simply puzzling that all these states corresponding to identical initial data in the bulk have very different semi-classical preparations. Of course, we could take any semi-classical state and remove the contribution of one energy eigenstate. That will also have an overlap close to one compared to the original state, but in no sense does the amputated state have a different semi-classical preparation. It is a small microscopic alteration of the state. Our states here have semi-classical preparations which are very different, and are drastically different at the microscopic level. They are simply not different enough to alter the overlap.

We should also note that we should really separate the moduli of the original genus-2 surface defining the half genus-2 state in two: the modulus of the handle and the modulus of the cylinder. Given a fixed modulus of the cylinder, there is a unique choice of $\beta$ for which this overlap gives one. In turn, this uniquely fixes $\tilde{\beta}$. On the other hand, the modulus of the handles can be tuned at will here, without affecting the choice of $\beta$ or the leading order value of the overlap. This is the CFT incarnation of what we found in the bulk: there are many different choices of wormhole states that all have large overlap with BTZ.

\subsection{Comments on the $U(1)$ symmetry.}

There is one aspect of the wormhole state that is worth discussing in more detail and is related to symmetry. The thermofield double state is rotationally invariant, this means that it is annihilated by the sum of the spin operators\footnote{Note the sign difference between the sum of spin operators and the operator that annihilates the thermofield double state $H_L-H_R$. This is due to the CPT operation that takes place between the left and right states.}
\be
(J_L + J_R) \ket{\text{TFD}}=0 \,.
\ee
This is also manifest in the preparation of the state, which is done on a cylinder that is rotationally invariant. This symmetry extends into the bulk, giving a killing vector of the spacetime. For the wormhole state, there is no such symmetry in its preparation. This means we expect
\be
(J_L + J_R) \ket{\psi_{\text{WH}}}\neq0 \,.
\ee
When we stare at the definition of the wormhole state, we see from \rref{ansatzwh} that the CPT nature of the left-right pairing guarantees that at the level of the primaries, the operator $(J_L + J_R)$ annihilates the state. But this happens only at the level of the primaries. For the descendants, we will get non-zero contributions confirming indeed that $(J_L + J_R)$ does not annihilate the state. 

There is a version of the this happening in the bulk: while the preparation of the state does not preserve a $U(1)$ rotation symmetry, the initial data actually does preserve the symmetry. This suggests that the expectation value of $(J_L + J_R)$ vanishes in the semi-classical limit, and that its variance is small, and only visible at the quantum level. This is perfectly compatible with the state being annihilated by $(J_L + J_R)$ at the level of primaries, but not at the level of descendants. It would be interesting to see if there were a deeper connection between initial data, symmetries, and primary vs descendants in the dual CFT.

\subsection{A Lorentzian version of the factorization puzzle}

Before concluding, we will discuss a Lorentzian incarnation of the factorization puzzle \cite{Maldacena:2004rf}. The factorization puzzle was originally  formulated in the Euclidean context: the product of Euclidean CFT partition functions on disconnected manifolds does not factorize when computed in gravity. Here, we will discuss a Lorentzian version of the same paradox, which to the best of our knowledge has not appeared in the literature. We start by considering the product state of two half genus-2 states as in \rref{prodstate}. This state has two important properties: it is a generalized Hartle-Hawking state, meaning it is a semi-classical state in gravity with a clear preparation in the CFT. And it is a factorized state meaning that for the reduced density matrix
\be
\rho_L = \Tr_R \left[ \ket{\psi_{\frac{1}{2}g=2}}_L \otimes \ket{\psi_{\frac{1}{2}g=2}}_R  \bra{\psi_{\frac{1}{2}g=2}}_L \otimes \bra{\psi_{\frac{1}{2}g=2}}_R  \right] \,,
\ee
we have
\be
S_L =0 \,.
\ee
The state is completely factorized, and there is no correlation between the left and right copies. However, if we try to compute some connected correlation function between the left and right copies, we will find a non-zero answer
\be
\bra{\psi_{\frac{1}{2}g=2}}_L \otimes \bra{\psi_{\frac{1}{2}g=2}}_R  O_L O_R  \ket{\psi_{\frac{1}{2}g=2}}_L \otimes \ket{\psi_{\frac{1}{2}g=2}}_R \Big|_c \neq 0  \,.
\ee
This answer will comes from the genus-2 wormhole, which is an on-shell saddle point that contributes to the boundary conditions specified above. It will be exponentially small in the central charge, due to the fact that the wormhole action is exponentially suppressed compared to the disconnected handlebody contributions, but it will be non-zero. This is in sharp tension with the manifestly factorized property of the quantum state.

Naturally, this is not profoundly different than the original factorization puzzle, it simply correspond to doing a slicing of the original puzzle and think about the Hilbert space interpretation of it. It would be interesting to have a similar construction for the thermofield double state. One would start with a pure state on one CFT, of the form
\be \label{onesidedtfd}
\ket{\psi}= \sum_{n} e^{-\beta E_n/4} \ket{n} \,.
\ee
Now we can construct the direct tensor product of two such states, which again is manifestly factorized. We see that the thermofield double state is the diagonal projection of the two states, which would make it entangled, and is dual the bulk black hole with a connected Einstein-Rosen bridge. So this geometry would give a non-zero contribution to
\be
\bra{\psi}_L \otimes  \bra{\psi}_R O_L O_R\ket{\psi}_L \otimes  \ket{\psi}_R \Big|_c \,.
\ee
The advantage of the genus-2 example that we gave above, is that every single component, including the disconnected states, are prepared semi-classically with the Euclidean path integral, while it is not clear how to have a nice Hartle-Hawking like preparation of \rref{onesidedtfd}.

\section{Discussion \label{sec:discussion}}

In this paper, we analyzed the state obtained by slicing a genus-2 Euclidean wormhole along a time-reflection symmetric slice. We saw that the gravitational initial data was the same as that of the BTZ black hole, hinting at the fact that the two states should be close in the semi-classical limit. We argued that this was indeed the case, by computing the overlap of the two states, which could be done by a continuous cutting and gluing procedure. Finally, we conjectured a microscopic definition of the wormhole state, by doing a diagonal projection on the direct product of two half genus-2 states, resulting in an entangled state. We checked this conjecture at the level of the overlap of the state with itself, and its overlap with BTZ, finding agreement with the bulk calculations. We conclude with some open questions.

\subsection*{More general slicings}

A natural question is whether more general slicings than the one considered in this paper can also be studied. There are many cuts through genus-2 surfaces that lie on $\mathbb{Z}_2$ symmetric slices. In general, they can lead to a Hilbert space defined on multiple circles. The simplest example of that type would be to slice a genus-2 surface through the three tubes of the sunset channel (rather than the one tube of the dumbbell). We believe that similar considerations as the ones described in this paper can be applied there, and it appears possible to write down microscopic version of those states as well. It would most interesting to find some wormhole slicing that on the symmetric slice, gives rise to a multi-boundary wormhole \cite{Balasubramanian:2014hda}.

\subsection*{Including one-loop determinants}

It would also be very interesting to include one-loop determinants in the calculation of the overlaps. One-loop determinants are multiplicative, so we expect
\be
\braket{\text{TFD}|\psi_{\text{WH}}} = f(\beta) \,,
\ee
for some order one function (in terms of $c$) of the moduli. The cutting and gluing procedure of the bulk does not fix the ratios of determinants to be one, as the 1-loop determinants are not local functions of the geometry, but rather care about global aspects of the spacetime. It would be interesting to study how these 1-loop factors depend on the moduli, and how far they can push the overlaps from unity. Since all these 1-loop determinants capture the fluctuations of quantum fields on a given spacetime, it is natural to suspect that the different wormhole states are spanning the basis of quantum states on top of the Einstein-Rosen initial data. It would be interesting to have a better understanding of these quantum overlaps, and see if any new puzzles appear.

\section*{Acknowledgements}
It is a pleasure to thank Stefano Antonini, Jan de Boer, Luca Iliesiu, Diego Liska, Kyriakos Papadodimas, Martin Sasieta and Ben Withers for fruitful discussions. I am also thankful to Daniel Jafferis and Gabor Sarosi for collaboration at an earlier stage of this project. I would also like to thank Veronica Hubeny and Mukund Rangamani for comments on the draft, as well as the Centro de Ciencias de Benasque Pedro Pascual for hospitality during the completion of this work.

\bibliographystyle{ytphys}
\bibliography{ref}

\providecommand{\href}[2]{#2}\begingroup\raggedright\begin{thebibliography}{10}

\bibitem{Maldacena:2004rf}
J.~M. Maldacena and L.~Maoz, ``{Wormholes in AdS},''
  \href{http://dx.doi.org/10.1088/1126-6708/2004/02/053}{{\em JHEP} {\bfseries
  02} (2004) 053}, \href{http://arxiv.org/abs/hep-th/0401024}{{\ttfamily
  arXiv:hep-th/0401024}}.

\bibitem{Maldacena:2001kr}
J.~M. Maldacena, ``{Eternal black holes in anti-de Sitter},''
  \href{http://dx.doi.org/10.1088/1126-6708/2003/04/021}{{\em JHEP} {\bfseries
  0304} (2003) 021},
\href{http://arxiv.org/abs/hep-th/0106112}{{\ttfamily arXiv:hep-th/0106112
  [hep-th]}}.

\bibitem{Saad:2018bqo}
P.~Saad, S.~H. Shenker, and D.~Stanford, ``{A semiclassical ramp in SYK and in
  gravity},'' \href{http://arxiv.org/abs/1806.06840}{{\ttfamily
  arXiv:1806.06840 [hep-th]}}.

\bibitem{Saad:2019lba}
P.~Saad, S.~H. Shenker, and D.~Stanford, ``{JT gravity as a matrix integral},''
  \href{http://arxiv.org/abs/1903.11115}{{\ttfamily arXiv:1903.11115
  [hep-th]}}.

\bibitem{Antonini:2022blk}
S.~Antonini, P.~Simidzija, B.~Swingle, and M.~Van~Raamsdonk, ``{Cosmology from
  the vacuum},'' \href{http://dx.doi.org/10.1088/1361-6382/ad1d46}{{\em Class.
  Quant. Grav.} {\bfseries 41} no.~4, (2024) 045008},
  \href{http://arxiv.org/abs/2203.11220}{{\ttfamily arXiv:2203.11220
  [hep-th]}}.

\bibitem{Antonini:2022ptt}
S.~Antonini, P.~Simidzija, B.~Swingle, and M.~Van~Raamsdonk, ``{Accelerating
  Cosmology from a Holographic Wormhole},''
  \href{http://dx.doi.org/10.1103/PhysRevLett.130.221601}{{\em Phys. Rev.
  Lett.} {\bfseries 130} no.~22, (2023) 221601},
  \href{http://arxiv.org/abs/2206.14821}{{\ttfamily arXiv:2206.14821
  [hep-th]}}.

\bibitem{Betzios:2024oli}
P.~Betzios and O.~Papadoulaki, ``{Inflationary Cosmology from Anti-de Sitter
  Wormholes},'' \href{http://dx.doi.org/10.1103/PhysRevLett.133.021501}{{\em
  Phys. Rev. Lett.} {\bfseries 133} no.~2, (2024) 021501},
  \href{http://arxiv.org/abs/2403.17046}{{\ttfamily arXiv:2403.17046
  [hep-th]}}.

\bibitem{Hubeny:2007xt}
V.~E. Hubeny, M.~Rangamani, and T.~Takayanagi, ``{A Covariant holographic
  entanglement entropy proposal},''
  \href{http://dx.doi.org/10.1088/1126-6708/2007/07/062}{{\em JHEP} {\bfseries
  07} (2007) 062},
\href{http://arxiv.org/abs/0705.0016}{{\ttfamily arXiv:0705.0016 [hep-th]}}.

\bibitem{Skenderis:2008dh}
K.~Skenderis and B.~C. van Rees, ``{Real-time gauge/gravity duality},''
  \href{http://dx.doi.org/10.1103/PhysRevLett.101.081601}{{\em Phys. Rev.
  Lett.} {\bfseries 101} (2008) 081601},
  \href{http://arxiv.org/abs/0805.0150}{{\ttfamily arXiv:0805.0150 [hep-th]}}.

\bibitem{Botta-Cantcheff:2015sav}
M.~Botta-Cantcheff, P.~Martinez, and G.~A. Silva, ``{On excited states in
  real-time AdS/CFT},'' \href{http://dx.doi.org/10.1007/JHEP02(2016)171}{{\em
  JHEP} {\bfseries 02} (2016) 171},
\href{http://arxiv.org/abs/1512.07850}{{\ttfamily arXiv:1512.07850 [hep-th]}}.

\bibitem{Marolf:2017kvq}
D.~Marolf, O.~Parrikar, C.~Rabideau, A.~I. Rad, and M.~Van~Raamsdonk, ``{From
  Euclidean Sources to Lorentzian Spacetimes in Holographic Conformal Field
  Theories},''
\href{http://arxiv.org/abs/1709.10101}{{\ttfamily arXiv:1709.10101 [hep-th]}}.

\bibitem{Belin:2018fxe}
A.~Belin, A.~Lewkowycz, and G.~Sarosi, ``{The boundary dual of the bulk
  symplectic form},''
  \href{http://dx.doi.org/10.1016/j.physletb.2018.10.071}{{\em Phys. Lett.}
  {\bfseries B789} (2019) 71--75},
\href{http://arxiv.org/abs/1806.10144}{{\ttfamily arXiv:1806.10144 [hep-th]}}.

\bibitem{Anderson:2020vwi}
L.~Anderson, O.~Parrikar, and R.~M. Soni, ``{Islands with gravitating baths:
  towards ER = EPR},'' \href{http://dx.doi.org/10.1007/JHEP10(2021)226}{{\em
  JHEP} {\bfseries 21} (2020) 226},
  \href{http://arxiv.org/abs/2103.14746}{{\ttfamily arXiv:2103.14746
  [hep-th]}}.

\bibitem{Maldacena:2018lmt}
J.~Maldacena and X.-L. Qi, ``{Eternal traversable wormhole},''
  \href{http://arxiv.org/abs/1804.00491}{{\ttfamily arXiv:1804.00491
  [hep-th]}}.

\bibitem{Belin:2020zjb}
A.~Belin and B.~Withers, ``{From sources to initial data and back again: on
  bulk singularities in Euclidean AdS/CFT},''
  \href{http://dx.doi.org/10.1007/JHEP12(2020)185}{{\em JHEP} {\bfseries 12}
  (2020) 185}, \href{http://arxiv.org/abs/2007.10344}{{\ttfamily
  arXiv:2007.10344 [hep-th]}}.

\bibitem{PhysRevLett.28.1082}
J.~W. York, ``Role of conformal three-geometry in the dynamics of
  gravitation,''
  \href{https://link.aps.org/doi/10.1103/PhysRevLett.28.1082}{{\em Phys. Rev.
  Lett.} {\bfseries 28} (Apr, 1972) 1082--1085}.

\bibitem{Belin:2018bpg}
A.~Belin, A.~Lewkowycz, and G.~Sarosi, ``{Complexity and the bulk volume, a new
  York time story},'' \href{http://dx.doi.org/10.1007/JHEP03(2019)044}{{\em
  JHEP} {\bfseries 03} (2019) 044},
\href{http://arxiv.org/abs/1811.03097}{{\ttfamily arXiv:1811.03097 [hep-th]}}.

\bibitem{Anderson:2001pf}
M.~T. Anderson, ``{Einstein metrics with prescribed conformal infinity on 4
  manifolds},'' \href{http://arxiv.org/abs/math/0105243}{{\ttfamily
  arXiv:math/0105243}}.

\bibitem{Hickling:2015tza}
A.~Hickling and T.~Wiseman, ``{Vacuum energy is non-positive for (2 +
  1)-dimensional holographic CFTs},''
  \href{http://dx.doi.org/10.1088/0264-9381/33/4/045009}{{\em Class. Quant.
  Grav.} {\bfseries 33} no.~4, (2016) 045009},
  \href{http://arxiv.org/abs/1508.04460}{{\ttfamily arXiv:1508.04460
  [hep-th]}}.

\bibitem{Fischetti:2016vfq}
S.~Fischetti, A.~Hickling, and T.~Wiseman, ``{Bounds on the local energy
  density of holographic CFTs from bulk geometry},''
  \href{http://dx.doi.org/10.1088/0264-9381/33/22/225003}{{\em Class. Quant.
  Grav.} {\bfseries 33} no.~22, (2016) 225003},
  \href{http://arxiv.org/abs/1605.00007}{{\ttfamily arXiv:1605.00007
  [hep-th]}}.

\bibitem{Jafferis:2017tiu}
D.~L. Jafferis, ``{Bulk reconstruction and the Hartle-Hawking wavefunction},''
  \href{http://arxiv.org/abs/1703.01519}{{\ttfamily arXiv:1703.01519
  [hep-th]}}.

\bibitem{Faulkner:2017tkh}
T.~Faulkner, F.~M. Haehl, E.~Hijano, O.~Parrikar, C.~Rabideau, and
  M.~Van~Raamsdonk, ``{Nonlinear Gravity from Entanglement in Conformal Field
  Theories},'' \href{http://dx.doi.org/10.1007/JHEP08(2017)057}{{\em JHEP}
  {\bfseries 08} (2017) 057}, \href{http://arxiv.org/abs/1705.03026}{{\ttfamily
  arXiv:1705.03026 [hep-th]}}.

\bibitem{Bahiru:2022oas}
E.~Bahiru, A.~Belin, K.~Papadodimas, G.~Sarosi, and N.~Vardian,
  ``{State-dressed local operators in the AdS/CFT correspondence},''
  \href{http://dx.doi.org/10.1103/PhysRevD.108.086035}{{\em Phys. Rev. D}
  {\bfseries 108} no.~8, (2023) 086035},
  \href{http://arxiv.org/abs/2209.06845}{{\ttfamily arXiv:2209.06845
  [hep-th]}}.

\bibitem{Bahiru:2023zlc}
E.~Bahiru, A.~Belin, K.~Papadodimas, G.~Sarosi, and N.~Vardian, ``{Holography
  and localization of information in quantum gravity},''
  \href{http://dx.doi.org/10.1007/JHEP05(2024)261}{{\em JHEP} {\bfseries 05}
  (2024) 261}, \href{http://arxiv.org/abs/2301.08753}{{\ttfamily
  arXiv:2301.08753 [hep-th]}}.

\bibitem{Haehl:2019fjz}
F.~M. Haehl, E.~Mintun, J.~Pollack, A.~J. Speranza, and M.~Van~Raamsdonk,
  ``{Nonlocal multi-trace sources and bulk entanglement in holographic
  conformal field theories},''
  \href{http://dx.doi.org/10.1007/JHEP06(2019)005}{{\em JHEP} {\bfseries 06}
  (2019) 005}, \href{http://arxiv.org/abs/1904.01584}{{\ttfamily
  arXiv:1904.01584 [hep-th]}}.

\bibitem{Belin:2020hea}
A.~Belin and J.~de~Boer, ``{Random Statistics of OPE Coefficients and Euclidean
  Wormholes},'' \href{http://arxiv.org/abs/2006.05499}{{\ttfamily
  arXiv:2006.05499 [hep-th]}}.

\bibitem{Chandra:2022bqq}
J.~Chandra, S.~Collier, T.~Hartman, and A.~Maloney, ``{Semiclassical 3D gravity
  as an average of large-c CFTs},''
  \href{http://dx.doi.org/10.1007/JHEP12(2022)069}{{\em JHEP} {\bfseries 12}
  (2022) 069}, \href{http://arxiv.org/abs/2203.06511}{{\ttfamily
  arXiv:2203.06511 [hep-th]}}.

\bibitem{Belin:2021ryy}
A.~Belin, J.~de~Boer, and D.~Liska, ``{Non-Gaussianities in the Statistical
  Distribution of Heavy OPE Coefficients and Wormholes},''
  \href{http://arxiv.org/abs/2110.14649}{{\ttfamily arXiv:2110.14649
  [hep-th]}}.

\bibitem{Collier:2023fwi}
S.~Collier, L.~Eberhardt, and M.~Zhang, ``{Solving 3d gravity with Virasoro
  TQFT},'' \href{http://dx.doi.org/10.21468/SciPostPhys.15.4.151}{{\em SciPost
  Phys.} {\bfseries 15} no.~4, (2023) 151},
  \href{http://arxiv.org/abs/2304.13650}{{\ttfamily arXiv:2304.13650
  [hep-th]}}.

\bibitem{Collier:2024mgv}
S.~Collier, L.~Eberhardt, and M.~Zhang, ``{3d gravity from Virasoro TQFT:
  Holography, wormholes and knots},''
  \href{http://dx.doi.org/10.21468/SciPostPhys.17.5.134}{{\em SciPost Phys.}
  {\bfseries 17} (2024) 134}, \href{http://arxiv.org/abs/2401.13900}{{\ttfamily
  arXiv:2401.13900 [hep-th]}}.

\bibitem{Belin:2023efa}
A.~Belin, J.~de~Boer, D.~L. Jafferis, P.~Nayak, and J.~Sonner, ``{Approximate
  CFTs and random tensor models},''
  \href{http://dx.doi.org/10.1007/JHEP09(2024)163}{{\em JHEP} {\bfseries 09}
  (2024) 163}, \href{http://arxiv.org/abs/2308.03829}{{\ttfamily
  arXiv:2308.03829 [hep-th]}}.

\bibitem{Jafferis:2024jkb}
D.~L. Jafferis, L.~Rozenberg, and G.~Wong, ``{3d Gravity as a random
  ensemble},'' \href{http://arxiv.org/abs/2407.02649}{{\ttfamily
  arXiv:2407.02649 [hep-th]}}.

\bibitem{deBoer:2024mqg}
J.~de~Boer, D.~Liska, and B.~Post, ``{Multiboundary wormholes and OPE
  statistics},'' \href{http://dx.doi.org/10.1007/JHEP10(2024)207}{{\em JHEP}
  {\bfseries 10} (2024) 207}, \href{http://arxiv.org/abs/2405.13111}{{\ttfamily
  arXiv:2405.13111 [hep-th]}}.

\bibitem{Hartman:2025ula}
T.~Hartman, ``{Triangulating quantum gravity in AdS$_3$},''
  \href{http://arxiv.org/abs/2507.12696}{{\ttfamily arXiv:2507.12696
  [hep-th]}}.

\bibitem{Chandra:2025fef}
J.~Chandra, ``{Statistics in 3d gravity from knots and links},''
  \href{http://arxiv.org/abs/2508.10864}{{\ttfamily arXiv:2508.10864
  [hep-th]}}.

\bibitem{Marolf:2017vsk}
D.~Marolf and J.~Wien, ``{The Torus Operator in Holography},''
  \href{http://dx.doi.org/10.1007/JHEP01(2018)105}{{\em JHEP} {\bfseries 01}
  (2018) 105}, \href{http://arxiv.org/abs/1708.03048}{{\ttfamily
  arXiv:1708.03048 [hep-th]}}.

\bibitem{Betzios:2021fnm}
P.~Betzios, E.~Kiritsis, and O.~Papadoulaki, ``{Interacting systems and
  wormholes},'' \href{http://dx.doi.org/10.1007/JHEP02(2022)126}{{\em JHEP}
  {\bfseries 02} (2022) 126}, \href{http://arxiv.org/abs/2110.14655}{{\ttfamily
  arXiv:2110.14655 [hep-th]}}.

\bibitem{Betzios:2023obs}
P.~Betzios and O.~Papadoulaki, ``{Wilson loops and wormholes},''
  \href{http://dx.doi.org/10.1007/JHEP03(2024)066}{{\em JHEP} {\bfseries 03}
  (2024) 066}, \href{http://arxiv.org/abs/2311.09289}{{\ttfamily
  arXiv:2311.09289 [hep-th]}}.

\bibitem{Maxfield:2016mwh}
H.~Maxfield, S.~Ross, and B.~Way, ``{Holographic partition functions and phases
  for higher genus Riemann surfaces},''
  \href{http://dx.doi.org/10.1088/0264-9381/33/12/125018}{{\em Class. Quant.
  Grav.} {\bfseries 33} no.~12, (2016) 125018},
\href{http://arxiv.org/abs/1601.00980}{{\ttfamily arXiv:1601.00980 [hep-th]}}.

\bibitem{Balasubramanian:2014hda}
V.~Balasubramanian, P.~Hayden, A.~Maloney, D.~Marolf, and S.~F. Ross,
  ``{Multiboundary Wormholes and Holographic Entanglement},''
  \href{http://dx.doi.org/10.1088/0264-9381/31/18/185015}{{\em Class. Quant.
  Grav.} {\bfseries 31} (2014) 185015},
  \href{http://arxiv.org/abs/1406.2663}{{\ttfamily arXiv:1406.2663 [hep-th]}}.

\end{thebibliography}\endgroup

\end{document}